\documentclass[twocolumn]{article}

\usepackage{cite} 
\usepackage{amsmath,amssymb,amsfonts}
\usepackage{algorithmic}
\usepackage{textcomp}

\usepackage[utf8]{inputenc} 
\usepackage{xcolor} 
\usepackage[colorlinks=true,urlcolor=urlblue,linkcolor=black,citecolor=black,breaklinks=true,bookmarks=false]{hyperref}
\usepackage{graphicx,epstopdf,float} 
\usepackage{authblk} 
\usepackage[justification=justified,singlelinecheck=false,font={footnotesize}]{caption} 
\usepackage[squaren]{SIunits} 

\newcommand{\bvec}[1]{ \mathbf{#1} }       
\newcommand{\uvec}[1]{ \mathbf{\hat{#1}} } 

\begin{document}
	
\title{
	Center of mass acceleration in coupled nanowaveguides due to transverse optical beating force
}

\author{Thales F. D. Fernandes, Cauê M. K. C. Carvalho, Paulo S. Soares Guimarães, Bernardo R. A. Neves, and Pierre-Louis de Assis
	\thanks{
		The authors acknowledge the funding by FAPEMIG, CNPq and CAPES.
		P.-L. de Assis was supported by CAPES Young Talents Fellowship Project number 88887.059630/2014-00.
		This work is part of CNPq project 457972/2014-9.		
	}
	\thanks{
		Thales F. D. Fernandes, Paulo S. Soares Guimarães, and Bernardo R. A. Neves are with the Physics Department, Federal University of Minas Gerais (UFMG), Minas Gerais, Brazil (e-mail: thalesfernandes@fisica.ufmg.br, pssg@fisica.ufmg.br, bernardo@fisica.ufmg.br)
	}
	\thanks{
		Cauê M. K. C. Carvalho and Pierre-Louis de Assis are with the Department of Applied Physics, ``Gleb Wataghin'' Institute of Physics, University of Campinas – UNICAMP, 13083-859, Campinas, São Paulo, Brazil (e-mail: kersulmt@ifi.unicamp.br and plouis@ifi.unicamp.br)
	}
}

\maketitle

\begin{abstract}
    Eigenmode optical forces arising in symmetrically coupled waveguides have opposite sign on opposite waveguides and thus can deform the waveguides by changing their relative separation, but cannot change any other degree of freedom on their own.
	It would be extremely desirable to have a way to act on the center of mass of such a system.
    In this work we show that it is possible to do so by injecting a superposition of eigenmodes that are degenerate in frequency and have opposite parity along the desired direction, resulting in beating forces that have the same sign on opposite waveguides and therefore act on the center of mass.
	We have used both the Maxwell Stress Tensor formalism and the induced dipole force equation to numerically calculate this transverse beating force and have found its magnitude to be comparable to the eigenmode forces. 
	We also show that the longitudinal variation caused by the spatial beating pattern on the time-averaged quantities used in the calculations must be taken into account in order to properly employ the divergence theorem and obtain the correct magnitudes.
	We then propose a heuristic model that shows good quantitative agreement with the numerical results and may be used as a prototyping tool for accurate and fast computation without relying on expensive numerical computation.
\end{abstract}

\section{Introduction}\label{sec:introduction}

Photonic integrated circuits (PICs) have steadily become a key area of technological development in the past decades, both for classical~\cite{Koch1991,Nagarajan2005} and quantum information processing~\cite{Sahin2013,Foster2006}, offering the possibility of larger processing speeds for classical bits and integration of a large quantity of qubits.
Two important features of PICs are the transverse confinement of the electromagnetic field, which constrains propagating waves into a discrete set of eigenmodes, and the evanescent fields-mediated interaction between dielectric waveguides~\cite{Okamoto2005}. 

When brought sufficiently close, two waveguides can interact evanescently forming a coupled-waveguides system (CWS) that has its own set of eigenmodes which may be symmetric or anti-symmetric with respect to the midpoint between waveguides.
The difference of propagation constants between eigenmdodes means that if the input is not an eigenmode the electromagnetic energy will shift from one waveguide to the other along the propagation direction forming a steady state beating pattern, this being the operating principle for directional couplers~\cite{Somekh1973} and beam splitters~\cite{Heaton1992} in PICs. 

Previous studies~\cite{Povinelli2005b} have shown that the dependence of propagation constants with waveguide separation gives rise to attractive and repulsive forces between waveguides, which depend mostly on the parity of the eigenmode in question.
The emergence of such optical forces has been thoroughly studied in the past decade, with both attractive and repulsive regimes having been observed~\cite{Li2009}.
These forces allow for a wide array of optomechanical effects to be built into PICs, such as the dynamical alignment of elements on a chip~\cite{Wiederhecker2009a,Li2008,Mizrahi2009}, the broadband modulation of light~\cite{Guo2012} and, most strikingly, giant optomechanical Kerr nonlinearities~\cite{Ma2011} and proofs of concept for non-volatile optomechanical memories \cite{Bagheri2011}.

While the light-matter interactions resulting from the excitation of a CWS with a superposition of eigenmodes have been well studied from the point of view of Brillouin scattering, the community studying optical forces between waveguides in contexts other than photon-phonon scattering has neglected superpositions almost completely~\cite{Russell1991,Butsch2012,Rakich2012,Wolff2015}.
Although cross terms in expressions for the optical force are briefly mentioned in the literature, called beating force~\cite{Li2009} ($F_{b}$), they were not fully understood and have been neglected on the grounds of being small compared to the eigenmode forces~\cite{Li2009} or just not considered relevant~\cite{Roels2009}.

We show in this paper that this is not the case in general.
The cross terms of the force density behave in a manner that is entirely different from those due to individual eigenmodes, with a magnitude that is comparable to the eigenmodes force if the forces are properly calculated.
In Section II we present numerical calculations for the cross terms using two models: the standard Maxwell Stress Tensor (MST), and the Dipole Force (DF) induced on the dielectric material by an inhomogeneous electric field distribution.
Our study considered a superposition of monochromatic guided eigenmodes of a pair of suspended waveguides.
The eigenmodes were degenerate in frequency, with different propagation constants.
We have only briefly considered electrostriction (ES), with the purpose of comparing the effects of its cross terms to that of the ones due to the optical force, which is derived for a dielectric with no photoelasticity or any consideration about the elastic properties of the waveguides.
Calculations were made using only undeformed geometries and no coupling to phonons is considered.

Following the numerical results, in Section III we propose a heuristic model based on conservation of total momentum and the orthogonality of the optical eigenmodes.
As mentioned before, it is well known that in a CWS energy shifts from one waveguide to the other if the input is not an eigenmode, meaning that the transverse position of the center of energy of the propagating light changes along the propagation direction.
As a consequence, the Poynting vector has a transverse component that varies along the direction of propagation.
This variation gives rise to a beating force density, which behaves in very good agreement with the one calculated numerically in Section II.

We finish by comparing the predictions of our heuristic model to the numerical results and discuss them in the context of distinguishing between the different force laws proposed for dielectrics, compare the magnitude of this radiation pressure force to the electrostrictive cross terms and propose possible applications of this force in CWS geometries optimised for obtaining large displacements with small input optical power.

\section{Numerical simulation}

In our work we consider a pair of silicon waveguides, with a refractive index $n=3.45$ at a wavelength $\lambda=1550~\mathrm{nm}$.
The geometry is of square cross-sections with dimensions $a=310~\mathrm{nm}$ and a horizontal separation $d$ that is swept during simulation with light propagating along the $z$-direction and we choose the relative phase of the eigenmodes to be in-phase.
Our numerical study is based in Finite Elements Method (FEM) simulations, where the total optical force was calculated by using the MST~\cite{Kemp2005,Rakich2011} formalism and as an induced DF~\cite{Yu2011,Zakharian2005,Mansuripur2008}.
All quantities are calculated in the frequency-domain, where we use a single frequency $\omega_0$ for both eigenmodes and all calculated quantities are time-averaged.

A superposition of eigenmodes can be written as
\begin{align}
	\bvec{E}(x,y,z) &= \sum_{\mu=1}^{2} A_\mu \bvec{E}_\mu(x,y) e^{i\beta_\mu z} , \label{Esuper} \\
	\bvec{H}(x,y,z) &= \sum_{\mu=1}^{2} A_\mu \bvec{H}_\mu(x,y) e^{i\beta_\mu z} \label{Hsuper},
\end{align}
where $\bvec{E}$ and $\bvec{H}$ are the total electric and magnetic field, respectively, $\beta_\mu$ is the propagation constant for the eigenmode $\bvec{E}_\mu,\bvec{H}_\mu$, and $|A_\mu|^2$ gives the power input of each eigenmode $\mu$.
For this work we used the first two TE-like modes of the CWS.

We first calculated the forces on waveguides from the MST, $\boldsymbol\sigma$, since it is the usual approach in CWS~\cite{Kemp2005,Rakich2011}:
\begin{align}
	{\boldsymbol\sigma} &= \frac{\epsilon_0}{2}\left( \operatorname{Re}\!\left[\bvec{E} \otimes \bvec{E}^*\right] - \frac{1}{2}\mathbb{I} \,\bvec{E}\cdot\bvec{E}^*\right)\\
	&+ 
	\frac{\mu_0}{2}\left( \operatorname{Re}\!\left[\bvec{H} \otimes \bvec{H}^*\right] - \frac{1}{2}\mathbb{I} \,\bvec{H}\cdot\bvec{H}^*\right), \nonumber
\end{align}
where $\mathbb{I}$ is the $3\times3$ identity matrix, and $\otimes$ is the outer-product between two vectors. The forces at a volume $V$ bounded by the waveguide boundary $\Omega$ are defined as
\begin{align}
	\bvec{F} &= \iiint_V \nabla \cdot {\boldsymbol\sigma} \, \mathrm{d}V =\iint_\Omega {\boldsymbol\sigma}^+ \cdot \uvec{n} \, \mathrm{d}S \label{FMST},
\end{align}
where we have used the divergence theorem.
Due to the field discontinuity, the field outside the waveguide, indicated as the superscript $+$, is used in the computation~\cite{Kemp2005}.

An important observation must be made with respect to the use of the divergence theorem in the case of superpositions.
When one wants to calculate the force density $\bvec{q}=\partial_z\bvec{F}$ along the propagation direction $z$, by making the volume $V$ in~\eqref{FMST} infinitesimally small in the $z$-direction, one needs to properly take into account the surface perpendicular to the propagation direction, since the stress tensor $\boldsymbol\sigma$ is not $z$-independent as it was in the case of eigenmodes.
A surface integral on the cross-section is therefore needed, besides the usual procedure of integrating over a curve on a cross section of the CWS.

The forces on the waveguides can also be calculated from the induced dipole force density $\bvec{f}$. Due to the discontinuity of the electric field introduced by the boundary between dielectric and vacuum, the induced DF can be split into a component acting on the bulk and one acting on the surface of the waveguides:
\begin{align}
	\bvec{f}_{\mathrm{bulk}} &= \frac{1}{4} \epsilon_0 \left( \epsilon-1\right) \nabla \!\left( \bvec{E} \cdot \bvec{E}^* \right), \label{DF1} \\
	\bvec{f}_{\mathrm{surf}} &= \frac{1}{4} \epsilon_0 \uvec{n}\cdot\left(\bvec{E}^+ - \bvec{E}^-\right)^*\left( \bvec{E}^+ - \bvec{E}^-\right), \label{DF2}
\end{align}
where the superscripts $-$ indicates the fields inside the material, $\uvec{n}$ is the outward normal to the waveguide boundary, and~$\varepsilon$ is the relative permittivity.

\begin{figure}[htbp]
	\centering
	\includegraphics[width=8cm]{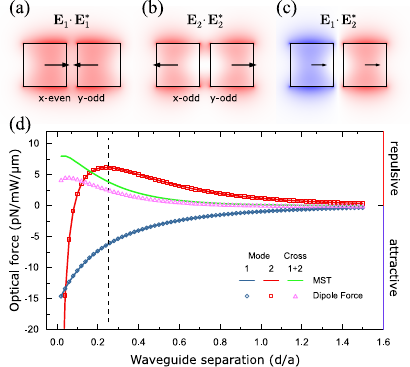}
	\caption{
		Results of FEM simulations for the CWS using the first two TE-like modes.
		The symmetric mode \textbf{(a)}, is $x$-even and $y$-odd, while the anti-symmetric mode \textbf{(b)} is $x$-odd and $y$-odd.
		In both cases the resulting force, indicated by a black arrow, depends of a square modulus and so is $x$-even, acting with opposite signs on each waveguide.
		For the beating force \textbf{(c)}, which depends on cross terms like $E_1^{}\cdot E_2^*$, the parity is $x$-odd, so that in both waveguides the force will act with the same sign.
		In each cross section, red indicates regions of positive $E_\mu^{}\cdot E_\nu^*$ and blue indicates regions where the product is negative.
		\textbf{(d)} The magnitudes for the attractive (blue/circles) and repulsive (red/squares) force agree remarkably well between calculations by MST (continuous lines) or DF (symbols).
		For the amplitude of the beating force (green/triangles), however, we see that the continuous line of the MST calculation does not coincide with the result from DF.
		The vertical dashed line indicates the separation $d/a\approx 0.26$ where the attractive and repulsive forces are equal in magnitude and should cancel each other.
	}
	\label{fig:mode-force}
\end{figure}

The results of the FEM simulation are shown in Fig.~\ref{fig:mode-force}, and we present both the known eigenmode forces and the amplitude of the beating force, since its magnitude varies sinusoidally along the direction of propagation, calculated taking the aforementioned precautions.

Fig.~\ref{fig:mode-force}(a,b) shows the electric field squared norm for eigenmodes $1$ and $2$ over a cross section of the CWS.
The symmetry of the given eigenmode is indicated as $x(y)$-even(odd) as appropriate.
Fig.~\ref{fig:mode-force}(c) shows the cross-term of the electric field squared norm, results in a profile which is odd in the $x$-plane.

Due to the opposite symmetries of the chosen modes, the cross-terms for both MST and DF will be $x$-odd, while eigenmode terms are $x$-even.
The odd parity about the $x$-plane results in a force that acts on both waveguides with the same sign, so that a cross section of the CWS will experience a non-zero net force.
This characteristic of the beating force makes it able to accelerate the center of mass of an adequately designed finite pair of waveguides.

Fig.~\ref{fig:mode-force}(d) shows the magnitude of the optical forces in the system, highlighting the separation $d/a\approx0.26$ where the attractive (blue) and repulsive (red) forces cancel each other and only the beating force (green) is felt.
The results using MST are shown as continuous lines and the DF by symbols, circles for the attractive force, squares for the repulsive force and triangles for the beating force.
It is possible to see that both formulations agree remarkably well among themselves for eigenmodes, hence eigenmode forces are unambiguous.
For $q_b$, however, we see that the formulations do not agree.

\section{Heuristic model} \label{sec:heuristic}

The aim of our model is to calculate the forces acting on a CWS when the pair is excited in a superposition of eigenmodes, without resorting to the MST formalism and in a way that provides more intuition about the momentum exchanges between light and matter in this system.
It does so by calculating the change in the linear momentum of the electromagnetic field as the center of energy of the system changes transversally along the direction of propagation, from one waveguide to the other.

We start with the (time averaged) Poynting vector $\bvec{S}$, which is defined as
\begin{align} \label{eq:Poynting}
	\bvec{S} &= \operatorname{Re} \left[ \bvec{E} \times \bvec{H}^* \right] / 2.
\end{align}
Substituting the superposition fields from~\eqref{Esuper} and~\eqref{Hsuper} into~\eqref{eq:Poynting}, we have:
\begin{align}
	\bvec{S} &= \frac{1}{2}
		\sum_{\mu=1}^{2}\sum_{\nu=1}^{2} \operatorname{Re}\left[
			A_\mu A^*_\nu
			\bvec{E}_\mu \times \bvec{H}^*_\nu
		\right] e^{i \Delta\beta_{\mu\nu}z}  \\
	&= \frac{1}{4} \sum_{\mu=1}^{2}\sum_{\nu=1}^{2} A_\mu A^*_\nu \left[ 
		\bvec{E}_\mu \times \bvec{H}^*_\nu + \bvec{E}_\nu^* \times \bvec{H}_\mu 
	\right] e^{i \Delta\beta_{\mu\nu}z} ,
\end{align}
where $\Delta\beta_{\mu\nu}=\beta_\mu-\beta_\nu$ and the labelling inversion on the last part is allowed due to the ranges of the sums over $\mu$ and $\nu$ being the same. 
By defining
\begin{align}
	\bvec{S}_{\mu\nu} &= \left( 
		\bvec{E}_\mu \times \bvec{H}_\nu^* + \bvec{E}_\nu^* \times \bvec{H}_\mu
	\right)/4, \label{Smn}
\end{align}
the Poynting vector can be written as:
\begin{align}
	\bvec{S} &= \sum_{\mu=1}^{2}\sum_{\nu=1}^{2}  A_\mu^{} A_\nu^* \bvec{S}_{\mu\nu} e^{i \Delta\beta_{\mu\nu}z} \\
	&= |A_1|^2 \bvec{S}_{11} + |A_2|^2 \bvec{S}_{22} + 2 \operatorname{Re}\left[ A_1^{} A_2^*\bvec{S}_{12} e^{i \Delta\beta_{\mu\nu}z} \right],
\end{align}
where we used the fact that $\bvec{S}_{\mu\nu}=\bvec{S}_{\nu\mu}^*$, which can be seen from~\eqref{Smn}.
The transverse component of $\bvec{S}_{\mu\nu}$ will be totally imaginary whereas the longitudinal one will be real, which can be shown from the eigenmode fixed phase, therefore they will show respectively a sine and cosine dependence with the propagation direction.

The cross terms of any quantity quadratic in the fields can be calculated in the same manner, and we see explicitly that only cross terms will be $z$-dependent.

Light carries momentum density $\bvec{g}$ that can be expressed, respectively, in the Abraham or the Minkowski definition~\cite{Yu2011}:
\begin{align}
	\bvec{g}_A &= \frac{1}{2c^2} \operatorname{Re} \left[ \bvec{E} \times \bvec{H^*} \right] = \frac{1}{c^2} \bvec{S} \label{eq:gA}, \\
	\bvec{g}_M &= \frac{1}{2} \operatorname{Re} \left[ \bvec{D} \times \bvec{B	^*} \right] = \frac{\epsilon}{c^2} \bvec{S}, \label{eq:gM}
\end{align}
where $\bvec{D}=\varepsilon_0\varepsilon\bvec{E}$ is the electric displacement field, $\bvec{B}=\mu_0\bvec{H}$ is the magnetic flux density, and we have assumed a non-magnetic, dispersionless and absorptionless medium.
Therefore, momentum density is directly proportional to the energy density flux $\bvec{S}$.

As mentioned before, energy will shift from one waveguide to the other.
Therefore its center of energy will also shift from one to another.
In electromagnetic systems, energy density~\cite{Jackson1998,Povinelli2005a} is defined as:
\begin{align}
	u &= \left( \bvec{E}\cdot\bvec{D}^* + \bvec{H}\cdot\bvec{B}^* \right) / 4
\end{align}
where the extra one-half factor comes from time-averaging. 

The center of energy~\cite{Kowalski2010} ${\boldsymbol\rho}_c$ is defined as:
\begin{align}
	{\boldsymbol\rho}_c(z) &= \frac{1}{U} \iint_\Omega {\boldsymbol\rho}u \, \mathrm{dS},
\end{align}
where ${\boldsymbol\rho}$ is the transverse vector in the $xy$-plane and $U$ is the energy in the cross-section.
The eigenmodes of symmetrical coupled-waveguides will have either even or odd transverse profiles.
Therefore the term $u_{11}$ or $u_{22}$ will always be even since it is the product of two even fields or two odd fields.
Since ${\boldsymbol\rho}$ is odd, a term of the form ${\boldsymbol\rho}u_{11}$ or ${\boldsymbol\rho}u_{22}$ will always be odd and have null integral.
On the other hand, the cross-term $u_{12}$ can be odd if the eigenmodes have different parity, similarly to Fig.~\ref{fig:mode-force}(c).
In this case ${\boldsymbol\rho}u_{12}$ will be even and have a non-null integral.
If the eigenmodes have the right kind of parity, it is then possible to have a center of energy that shifts from one waveguide to the other.

\begin{figure}[!h]
	\centerline{\includegraphics[width=\columnwidth]{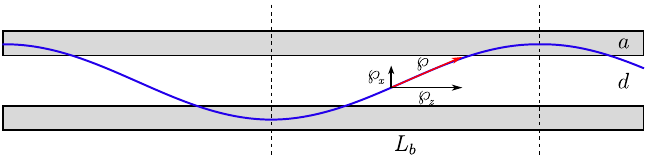}}
	\caption{
		Diagram of coupled waveguides exchanging energy.
		The blue sinusoidal line shows the light-path (center of energy), where $z$ is the propagation direction and $x$ is the transverse direction.
		The waveguides have width $a$, are separated by a distance $d$ and the energy exchange is characterized by $L_b$, the beating distance, which is the distance light couples totally from one waveguide to the other.
		${\boldsymbol\wp}$ is the momentum at the cross-section. 
	}
	\label{fig:wg-energy-center}
\end{figure}

If the waveguide separation $d$ is not too small relative to their width and the superposition is composed of equal amounts of each orthonormalized eigenmode ($A_1 = A_2$) 
we can approximate the center of energy as the distance from the center of the CWS to the center of one of the waveguides. 
Therefore, a reasonable assumption for the movement of the center of energy is
\begin{align}
	{\boldsymbol\rho}_c(z) &= \frac{d+a}{2}\uvec{x} \cos\left( \frac{\pi z}{L_b} \right),
\end{align}
where $x$ is the transverse direction and $L_b$ is the beating distance, which is the distance light couples totally from one waveguide to the other.
At $z=0$, light is totally on the top waveguide as shown in Fig.~\ref{fig:wg-energy-center}.
The momentum in the transverse cross-section is ${\boldsymbol\wp}$ and it is always tangent to the light-path.
Due to the orthogonality of eigenmodes, ${\boldsymbol\wp}_z$ will always be constant, irrespectively of $z$~\cite{Mahmoud1991,Chuang2009}.

While the center of energy is analogous to the center of mass, it should be kept in mind that the electromagnetic field is delocalized in the CWS.
Therefore, the trajectory obtained is not a literal one, in the sense of a point particle moving through space.

For the linear momentum to be tangent we must have:
\begin{align}
	\frac{\wp_x}{\wp_z} &= \frac{d}{dz}\rho_{xc}(z) =
	-\frac{d+a}{2} \frac{\pi}{L_b} \sin\left( \frac{\pi z}{L_b} \right), \label{eq:px}
\end{align}
which gives a relationship for the linear momentum transverse component on both waveguides.
We shall express the beating force as the force density $q_{xb}=\partial_z F_{xb}$ along the $z$-direction and it can be found from the variation of the linear momentum, $\bvec{q}_b=d\bvec{\boldsymbol\wp}/dt$, and the time derivative can be converted into a spatial derivative by knowledge of the propagation velocity $v$~\cite{Mansuripur2008}.
The beating force density is then defined as:
\begin{align}
	q_{xb} &= v \frac{d}{dz} \wp_x = 
	\wp_z \left(\frac{\pi}{L_b}\right)^2 \frac{d+a}{2} v \cos\left( \frac{\pi z}{L_b} \right).
\end{align}

This force acts on both waveguides, and since we are dealing with symmetric waveguides, the force on each waveguide will be half of this value, $q_{xb}^\mathrm{(T/B)}=q_{xb}/2$, where the superscript T/B indicates the top and bottom waveguide, respectively.
It is important to point out that this force is capable of accelerating the center of mass of the system, since for all cross sections it has the same sign on both waveguides.

The beating force calculated by the heuristic model presented above depends on how the momentum of light in the CWS is calculated, $\wp_z$.
This means that choosing to use the Abraham or Minkowski formulations will entail different quantitative predictions for $\bvec{q}_b$. 

\section{Discussion} \label{Discussion}

Having ascertained that the two sets of equations used to calculate optical forces from our FEM simulations are consistent with each other for eigenmodes we proceeded to use the numerical results for cross terms as validation for our heuristic model.

\begin{figure}[htbp]
	\centering
	\includegraphics[width=8cm]{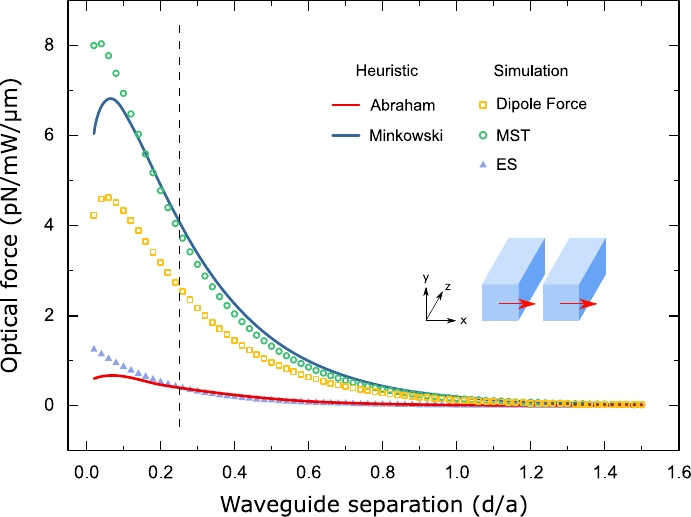}
	\caption{
		Beating force as a function of the adimensional waveguide separation, $d/a$. Five derivations are presented: calculation via momentum transfer, assuming both Abraham (red line) and Minkowski (blue line) formulations, and calculation using MST (green circle), DF (yellow square), and ES (blue triangles), based solely on numerical FEM simulations.
	}
	\label{fig:beat-force}
\end{figure}

The beating force, shown in Fig.~\ref{fig:beat-force}, was calculated from FEM data, using both MST and DF formulations.
Since electrostrictive effects are commonly observed in silicon nanowaveguides, we have also calculated the magnitude of that effect, shown in Fig.~\ref{fig:beat-force} as blue triangles.
It is clear from the results that the contribution of electrostriction is always much smaller than the beating force and therefore should not interfere in experimental observations of the beating force.
The forces shown were markedly different for each derivation.
For Abraham momentum the predicted force is very small, whereas for Minkowski momentum gives a much higher force, as expected from the presence of an $\epsilon\approx 10$ in~\eqref{eq:gM}.
The heuristic force calculated from Minkowski momentum agrees remarkably well with MST, even though the momentum associated with MST is Abraham one, whereas DF have a smaller force but with the same general behavior.
The difference in momentum depends on the system of interest, with Minkowski momentum associated to canonic momentum while Abraham to kinetic momentum~\cite{Barnett2010}.
This may suggest that in waveguide systems the correct momentum formulation is the Minkowski one.

\begin{figure}[htbp]
	\centering
	\includegraphics[width=8cm]{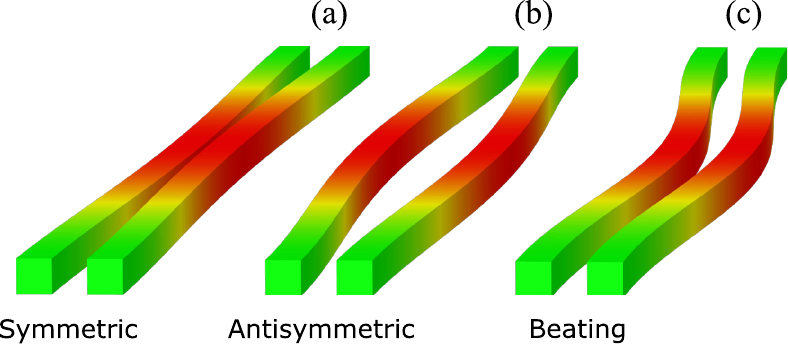}
	\caption{
		Mechanical displacement caused by the symmetric (\textbf{a}) and antisymmetric (\textbf{b}) eigenmode forces, compared to the beating force (\textbf{c}).
		Displacements shown are solutions of the Euler-Bernoulli equation for a pair of doubly clamped waveguides with a square cross-section.
		The scale of deformations has been exaggerated from actual values in order to better display the differences in the resulting deformations.
		It can be seen that, while eigenmode forces do not act on the center of mass of the system, but change the distance between waveguides, the opposite is true for the beating force.
		For the latter, both waveguides are accelerated in the same direction, resulting in a displacement of the center of mass, while the separation between waveguides does not change.	
	}
	\label{fig:disp}
\end{figure}

To compare the eigenmode forces with the beating force we chose to simulate its effect on a doubly-clamped CWS as demonstrated in Fig.~\ref{fig:disp}.
Fig.~\ref{fig:disp}(a) and Fig.~\ref{fig:disp}(b) show the deformation of CWS for symmetric and antisymmetric eigenmode forces, respectively, where the color map shows the stresses.
As stated, this force can only change the relative separation of both waveguides.
For a superposition of eigenmodes, Fig.~\ref{fig:disp}(c), the beating force can change the center of mass.
By a careful choice of geometry and/or controlling the amount of each eigenmode into the superposition, it is possible to control the deformation of each waveguide independently from one another, thus offering an additional degree of freedom in the mechanical manipulation of waveguides through optical forces. 

\section{Conclusions and outlook}

In this work we have shown that numerical calculations of the magnitude of the transverse beating force in coupled-waveguide systems must be done differently from the standard procedure used for eigenmodes. For this geometry, ignoring contributions to the force which are due to the variation of the MST along the direction of propagation leads to an under-estimation of the magnitude of $q_b$. In addition to that, we have shown that this force is neither attractive nor repulsive, so that it may accelerate the center of mass of an adequately designed pair of waveguides.

Motivated by this counterintuitive result, we proposed a heuristic model based on simples assumptions about the conservation of momentum in the total system, taking into consideration the transverse variation of the position of the center of energy of the electromagnetic field along the direction of propagation. This heuristic model was able to quantitatively predict the amplitude of $q_b$ with very good agreement to the results of the MST calculation up to a region of parameters where the initial assumptions fail. Interestingly, this good quantitative agreement is obtained when using the Minkowski formulation for the momentum of light in dielectrics, even though the MST is based on the Abraham momentum.

While we have presented our results for a pair of waveguides, they are more general and also apply for a single multi-mode waveguide. The choice of modes used in our calculations is also not specific and different mode superpositions will generate beating forces along the $y$ direction. By choosing a suitable superposition of modes as input, it should be possible to achieve independent manipulation of waveguides on both transverse directions, for instance in order to align photonic components between chips. Moreover, the fact that the beating force cannot always be neglected implies that the design of optomechanical devices based on optical forces between waveguides should take it into account whenever non-eigenmode excitation is expected.

\section*{Acknowledgment}
The authors would like to thank Pablo Saldanha, Thiago Alegre and Gustavo Wiederhecker for the fruitful discussions.


\end{document}